\documentclass[10pt,twocolumn]{article}
\usepackage[
  letterpaper,
  top=0.7in,
  bottom=0.8in,
  left=0.65in,
  right=0.65in
]{geometry}
\usepackage{graphicx}
\usepackage{amsmath, amssymb}
\usepackage{booktabs}
\usepackage{caption}
\usepackage[numbers,sort&compress]{natbib}
\usepackage{authblk}
\usepackage{enumitem}
\usepackage{microtype}
\usepackage{balance}
\usepackage{xurl}
\usepackage[hidelinks]{hyperref}
\usepackage{float}

\hypersetup{
    colorlinks=true,
    linkcolor=black,
    citecolor=black,
    urlcolor=blue
}
\hypersetup{
  pdftitle={Garbage Collection and Energy Consumption in Java: A Controlled Study Across Workloads and JDKs},
  pdfauthor={Rahil Sharma},
  pdfsubject={Garbage Collection, Energy Efficiency, Java Virtual Machine, Empirical Software Engineering},
  pdfkeywords={garbage collection, energy consumption, Java Virtual Machine, workload intensity, JDK distribution, RCBD}
}

\title{Garbage Collection and Energy Consumption in Java: A Controlled Study Across Workloads and JDKs}

\author{
Rahil Sharma\\
Vrije Universiteit Amsterdam\\
\texttt{r.sharma4@student.vu.nl}
}

\date{July 2026}

\begin{document}

\maketitle

\begin{abstract}
Garbage-collector selection is a low-effort configuration decision that can influence both the performance and energy consumption of Java applications. However, it remains unclear whether aggregate energy rankings among collectors generalise across heterogeneous applications, workload intensities, and JDK distributions. This study presents a controlled empirical evaluation of Serial, Parallel, and G1 garbage collection across three Java applications, three workload intensities, and two JDK distributions: OpenJDK and Oracle~JDK. Using EnergiBridge, processor-package energy consumption and execution time were measured and complementary energy--performance metrics derived. Across all evaluated configurations, Parallel recorded the lowest mean energy consumption (839.8~J), followed closely by Serial (857.6~J) and G1 (969.0~J), but an RCBD ANOVA did not establish a statistically reliable collector effect. Workload intensity, by contrast, was a significant driver of energy consumption, with heavy workloads consuming substantially more energy than light or medium workloads regardless of collector. Energy consumption showed a moderate positive association with execution time ($r = 0.33$), indicating that longer-running configurations tended toward higher energy use, though the relationship was far from proportional. No statistically significant energy difference was found between OpenJDK and Oracle~JDK. Overall, the results do not support a universally energy-optimal garbage collector; instead, workload intensity emerges as the more reliable lever for managing Java energy consumption, and collector selection should be treated as an application-specific tuning decision supported by measurement on the target system rather than aggregate rankings alone.
\end{abstract}

\section{Introduction}
\label{sec:introduction}

Energy consumption is becoming an important quality concern in software engineering alongside established performance objectives such as execution time, throughput, and latency. The global electricity consumption of the information and communication technology sector was estimated at approximately 1,000~TWh in 2023 \cite{ericsson2024}, corresponding to about 4\% of global electricity use \cite{statista2025}. These estimates have increased interest in green software engineering and in design practices that reduce the energy required to develop and operate software systems \cite{Fonseca2019}. The issue is particularly relevant in large-scale deployments, where modest savings at the level of an individual application may become consequential when replicated across many servers and repeated executions \cite{Pinto2017}. \\

\noindent Developers and operators nevertheless continue to select and tune software configurations primarily for performance, reliability, and operational predictability. Energy efficiency is rarely treated as a direct configuration objective. This creates an important practical question: can configuration choices that require little or no modification to application code provide meaningful energy savings without degrading performance? Garbage-collector selection is one such configuration choice for Java applications. Java remains widely used in enterprise, server-side, and data-intensive systems \cite{Shimchenko2024}. On the Java Virtual Machine (JVM), garbage collection automates memory management by identifying and reclaiming objects that are no longer reachable. This process reduces the burden of manual memory management, but it also consumes processor time, affects application execution, and may introduce pauses \cite{CarpenAmarie2015}. Different garbage collectors organize this work differently, which can influence runtime behaviour, resource utilization, and consequently the total energy consumed during execution. \\

\noindent Modern HotSpot JVMs provide several collectors designed for different performance objectives. Serial GC performs collection using a single thread and is generally intended for smaller heaps or environments with limited processing resources. Parallel GC performs collection using multiple threads and primarily targets application throughput. Garbage-First GC, commonly known as G1, divides the heap into regions and performs much of its work incrementally to balance throughput with predictable pause times \cite{Detlefs2004, OracleGCGuide2021}. G1 has been the default collector in HotSpot since Java~9. Together, these three collectors represent distinct and widely established approaches to garbage collection: single-threaded simplicity, parallel throughput optimization, and region-based pause-time management. These collectors were designed principally around performance and responsiveness rather than energy consumption. Their performance characteristics may therefore produce different energy outcomes, but a collector that performs well for one application may not behave similarly for another. Prior research has shown that changing the collector can reduce the energy consumption of some Java programs without necessarily sacrificing performance. In particular, Shimchenko \emph{et al.} found that alternatives to the default G1 collector can be more energy efficient for certain workloads \cite{Shimchenko2022}. This suggests that GC selection may serve as a low-effort energy optimization. However, it does not establish that one collector is consistently preferable across applications, workload intensities, or runtime distributions. Aggregate comparisons can also conceal substantial differences between applications whose execution times, allocation patterns, and memory demands vary considerably. \\

\noindent This paper examines how reliably garbage-collector selection influences the energy consumption of Java applications under heterogeneous experimental conditions. A controlled empirical study of Serial, Parallel, and G1 GC is conducted across three Java applications, three workload intensities, and two JDK distributions, OpenJDK and Oracle~JDK. For each evaluated configuration, processor-package energy consumption and execution time are measured using EnergiBridge, and complementary metrics describing power and energy-performance behaviour are derived. The study is guided by the following research question:

\begin{quote}
\emph{How much does garbage-collector selection affect the energy consumption and performance of Java applications across different applications, workload intensities, and JDK distributions?}
\end{quote}

\noindent The results distinguish between descriptive differences and statistically supported effects. Across the complete dataset, Parallel GC recorded the lowest mean energy consumption, followed closely by Serial GC, with G1 GC highest. However, the mixed-effects analysis did not establish a statistically reliable overall collector effect after accounting for variation between applications. Energy consumption showed a moderate positive association with execution time under the tested conditions, while no statistically significant energy difference was found between OpenJDK and Oracle~JDK. \\

This study makes three contributions. First, it provides a controlled comparison of three established garbage collectors across multiple Java applications, workload intensities, and JDK distributions. Second, it evaluates aggregate collector rankings together with an inferential analysis that accounts for repeated measurements and between-application heterogeneity. Third, it provides an empirical basis for treating garbage-collector selection as an application-specific tuning decision rather than assuming the existence of a universally energy-optimal collector. The results do not imply that GC tuning is irrelevant. Instead, they show why collector choices should be evaluated using measurements from the intended application, workload, and deployment environment.

\section{Related Work}\label{sec:related}

Research on software energy consumption has examined the influence of programming languages, runtime environments, and software design decisions. Castor provides a methodological introduction to estimating the energy footprint of software, including the challenges involved in connecting hardware-level measurements to software-level decisions \cite{castor2024}. Currie et al. translate these concerns into development practices and organizational principles for building more sustainable software systems \cite{currie2025}. Procaccianti et al. classify the broader green software engineering literature into measurement, tooling, and methodological contributions, while identifying a continuing need for controlled empirical validation \cite{procaccianti2016}. Hindle's \textit{Green Mining} approach further demonstrates that changes made during software development can be associated with changes in energy consumption \cite{hindle2012}. These studies establish the broader motivation for treating energy efficiency as a software engineering concern rather than solely a hardware problem. \\

\noindent Empirical studies of managed runtimes have shown that memory management can contribute materially to application energy consumption. Contreras and Martonosi measured the energy effects of garbage collection and reported that GC activity could account for up to 37\% of processor energy in their experiments \cite{contreras2006energy}. Their evaluation, however, considered collectors and synthetic workloads that predate current HotSpot defaults. Nou et al. later examined the effects of JVM garbage collectors on power consumption and showed that runtime configuration can produce measurable energy differences, although their evaluation covered a comparatively limited set of benchmarks and workloads \cite{nou2017}. Georges et al. did not focus specifically on energy, but their recommendations concerning JVM warm-up, repeated execution, and statistical analysis remain important for reliable Java experiments \cite{georges2007}. More recent studies have broadened the scope of runtime-level comparisons. Ournani et al. evaluated 52 JVM distributions from eight providers using the DaCapo and Renaissance benchmark suites \cite{ournani2021jvmenergy}. They found substantial variation in energy efficiency across JVM distributions and showed that parameters related to garbage collection and just-in-time compilation can influence the resulting energy profile. Their study provides broad coverage of JVM implementations, whereas the present work holds most runtime conditions constant and examines the joint effects of collector choice, workload intensity, application, and JDK distribution. Wang et al. introduced GEAR, a framework for comparing garbage collectors across Java, Go, and C\# through runtime-independent workload primitives \cite{11029923}. Their goal was to enable equivalent cross-runtime experiments and expose collector-specific scalability and efficiency behaviour. The present study instead remains within the Java ecosystem and investigates whether energy differences among three established HotSpot collectors support a consistent practical recommendation. \\

\noindent The closest prior work is the study by Shimchenko et al., which investigated whether changing OpenJDK's default G1 collector could reduce application energy consumption without modifying application code \cite{Shimchenko2022}. The authors profiled 35 Java applications drawn from four benchmark suites and evaluated Serial, Parallel, CMS, G1, ZGC, and Shenandoah under multiple heap sizes, Java versions, and collector-specific configurations. Using RAPL-based measurements, they found that replacing G1 could provide considerable energy savings for many applications. Their broad configuration search reported savings of up to 47\% when configurations were selected solely for energy consumption, and up to 40\% when performance degradation was limited to 5\%. They also investigated machine-learning methods for predicting energy-efficient configurations and thereby reducing the cost of profiling. Shimchenko's subsequent dissertation places this experiment within a broader programme of research on the energy efficiency of concurrent garbage collection \cite{Shimchenko2024}. The collector comparison established that fully concurrent collectors, particularly ZGC and Shenandoah, consumed more energy than less concurrent alternatives under the evaluated conditions. The later work therefore focused on improving concurrent collectors through dynamic heap sizing and scheduling GC work on energy-efficient or otherwise idle processor cores. The dissertation consequently demonstrates both that GC selection can provide substantial savings in some cases and that collector implementation and scheduling can be modified to improve energy efficiency. \\

\noindent The present study begins from the same practical premise that garbage-collector selection is an accessible configuration-level intervention, but it addresses a different question. Shimchenko et al. investigated the maximum savings obtainable by searching a broad space of collectors, heap sizes, Java versions, and runtime parameters. Their results establish that beneficial configurations exist, but they do not imply that a fixed collector ranking will generalize across applications or operating conditions. This study therefore examines whether the energy differences among Serial, Parallel, and G1 remain statistically reliable when workload intensity and JDK distribution are varied and between-application heterogeneity is taken into account. The aim is not to predict an optimal configuration or modify a collector. Instead, the study tests the robustness of the simpler practitioner-facing claim that replacing the default G1 collector with Serial or Parallel provides a generally reliable energy optimization. \\

Lengauer et al. provide complementary evidence concerning why such generalization may be difficult \cite{lengauer2017}. Their analysis of DaCapo, SPECjvm2008, and Scala benchmarks characterizes allocation rates, survivor ratios, object sizes, array density, GC frequency, and changes between continuous and burst allocation phases. The study shows that applications place markedly different demands on memory-management systems and recommends heap sizes of approximately three times the live set to permit representative GC behaviour. This heap-sizing principle is adopted in the present experimental design. Unlike Lengauer et al., however, the present study measures processor-package energy consumption and execution time directly. Their workload characterization helps explain why collector effects observed for one application may not transfer unchanged to another. Taken together, prior work establishes that garbage collection can affect energy consumption, that replacing the default collector can produce substantial savings for some applications, and that the magnitude of these effects depends on runtime configuration and application memory behaviour. To date, prior work has not specifically examined whether aggregate rankings among Serial, Parallel, and G1 remain statistically reliable in a controlled design that jointly varies application, workload intensity, and JDK distribution while accounting for repeated measurements and subject-level heterogeneity. This study addresses this narrower question. It complements earlier configuration-search and collector-optimization research by evaluating the limits of a universal collector recommendation under heterogeneous experimental conditions.

\section{Experimental Methodology}\label{sec:methodology}

\subsection{Experimental Overview and Design}

A controlled experiment was conducted to examine how garbage-collector selection affects the energy consumption and execution time of Java applications under different workload intensities and JDK distributions. Three HotSpot garbage collectors were evaluated: Serial, Parallel, and G1. Each collector was tested across light, medium, and heavy workload configurations using OpenJDK and Oracle~JDK. The experiment followed a Randomized Complete Block Design (RCBD), with Java subject acting as the blocking variable. Blocking was necessary because differences between applications were expected to account for a substantial proportion of the observed variation: the subjects differed in functionality, workload implementation, execution time, allocation patterns, and memory demand, and comparisons made without accounting for these differences could attribute application-level variation incorrectly to garbage-collector selection. Garbage collector was the primary experimental factor, with Serial, Parallel, and G1 as its three levels; workload intensity and JDK distribution were included as additional crossed factors. Within each block, the experimental conditions therefore combined three garbage collectors, three workload intensities, and two JDK distributions. Experimental conditions were replicated, and run order was randomized to reduce the potential influence of temporal effects, including background system activity and changes in hardware temperature. Because only three subjects were available as blocks, subject was treated as a fixed rather than random effect in the subsequent analysis (Section~\ref{sec:analysis}); descriptive collector rankings are interpreted alongside the inferential model rather than treated as evidence of a universally optimal collector.

\subsection{Goal Definition Using the GQM Framework}

The experiment was defined using the Goal-Question-Metric (GQM) framework to connect its practical objective to the research question and measured outcomes. Table~\ref{tab:gqm-framework} summarizes this relationship. \\

\begin{table}[t]
\caption{Goal-Question-Metric definition of the experiment}
\label{tab:gqm-framework}
\centering
\small
\renewcommand{\arraystretch}{1.2}
\begin{tabular}{p{0.20\linewidth} p{0.70\linewidth}}
\toprule
\textbf{Element} & \textbf{Definition} \\
\midrule
Goal &
Analyse the effects of garbage-collector selection on the energy consumption and execution time of Java applications. \\
\midrule
Purpose &
Evaluate whether differences among Serial, Parallel, and G1 support a consistent collector recommendation. \\
\midrule
Perspective &
Software developers, system operators, and empirical software-engineering researchers. \\
\midrule
Context &
Three Java applications executed at three workload intensities using OpenJDK and Oracle~JDK. \\
\midrule
Question &
How much does garbage-collector selection affect energy consumption and execution time across different applications, workload intensities, and JDK distributions? \\
\midrule
Metrics &
Processor-package energy consumption (J), execution time (s), and derived mean power (W). \\
\bottomrule
\end{tabular}
\end{table}

\noindent The GQM definition emphasizes that the objective is not merely to identify the collector with the lowest aggregate energy consumption. It also examines whether the observed differences remain consistent across applications and experimental conditions. Energy consumption and execution time are therefore considered jointly, while mean power is used to distinguish changes in power demand from changes caused primarily by execution duration. The design specified 216 experimental runs, corresponding to three applications, three garbage collectors, three workload intensities, two JDK distributions, and four repetitions. As described in Section~\ref{sec:analysis}, a subsequent data-integrity check led to the exclusion of results from the benchmark subjects originally planned alongside these applications; the RCBD reported here reflects only the retained, verified experimental structure.

\begin{figure}[H]
\centering
\includegraphics[width=0.5\textwidth]{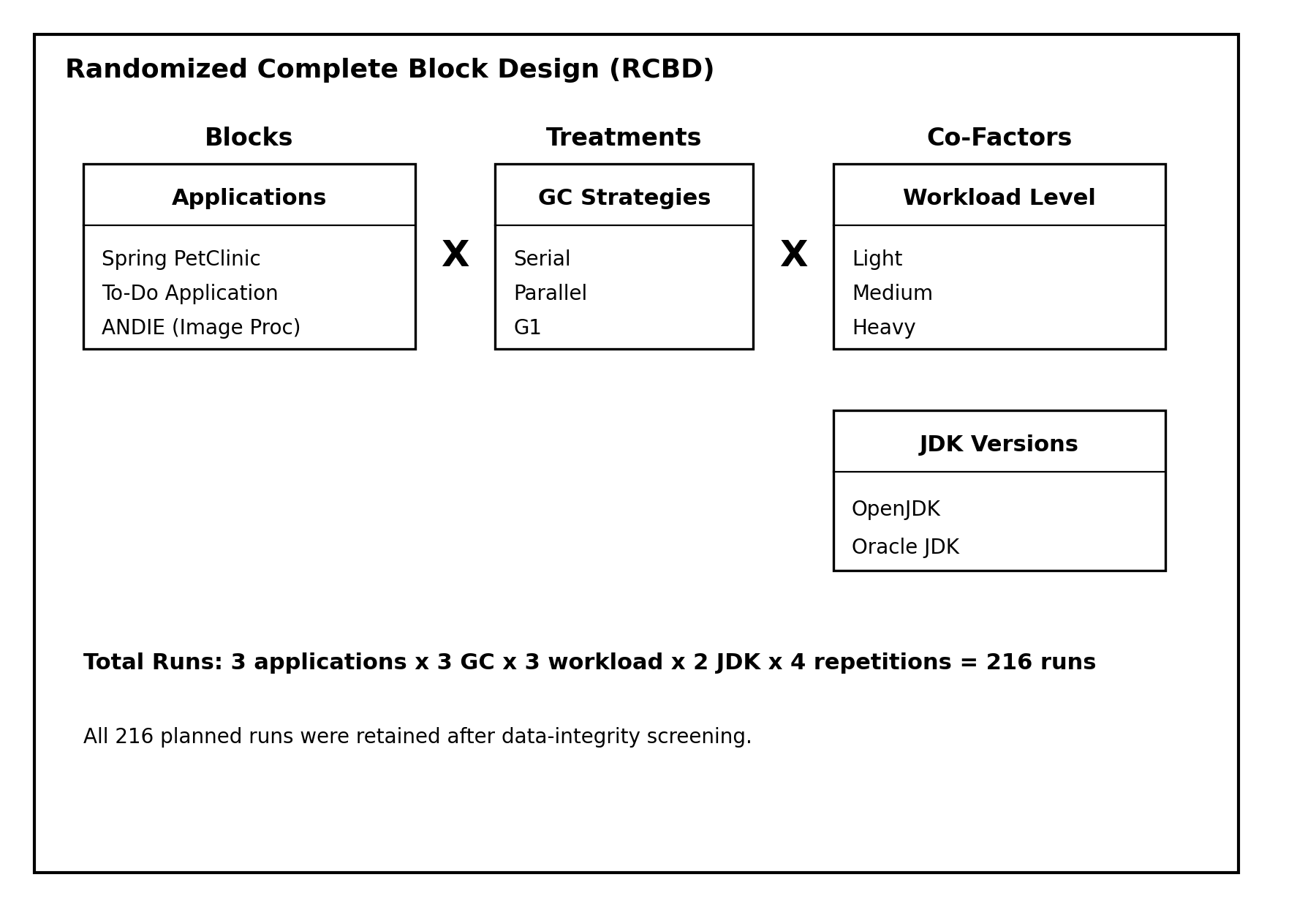}
\caption{Randomized Complete Block Design for the Java garbage-collection energy experiment, restricted to the three verified application subjects. All 216 planned runs were retained for analysis.}
\label{fig:rcbd-design}
\end{figure}

\subsection{Experimental Subjects}
\label{sec:subjects}

The experiment evaluated three Java applications: Spring PetClinic, a REST-based To-Do application, and the ANDIE image-processing tool. These subjects were selected to represent persistent, interaction-driven Java services rather than short-lived, terminating programs, so that collector behaviour could be examined under conditions closer to typical server-side deployment. The three applications differed in functionality, request or interaction patterns, and memory-allocation behaviour, which supported blocking on subject in the RCBD design described above.

\subsection{Workload Configuration}

Each Java subject was executed under light, medium, and heavy workload configurations. These levels increased the amount of work performed by the subject and, where applicable, the resulting allocation rate and heap pressure. Light configurations represented comparatively limited execution demands, medium configurations increased processing and memory allocation, and heavy configurations placed the greatest demand on the application and its memory-management system. The classification follows the broader observation that allocation rates, survivor ratios, object lifetimes, and garbage-collection frequency can differ considerably across Java workloads \cite{lengauer2017}. Because the evaluated subjects performed different tasks, the concrete parameters used to instantiate light, medium, and heavy workloads were defined separately for each subject. The workload levels should therefore be interpreted as ordered intensities within each subject rather than as identical quantities of work across all subjects.

\subsection{Experiment Orchestration and Execution}

The experiment was automated using the \texttt{ExperimentRunner} framework~\cite{experimentrunner}, deployed on a dedicated Raspberry~Pi~4, which initiated and monitored runs on a separate measurement laptop but did not execute the evaluated Java programs. This separation prevented the computational activity of the orchestration framework from contributing directly to the energy measurements collected on the laptop. For each run, ExperimentRunner selected the Java subject, workload intensity, JDK distribution, and garbage-collector configuration according to the randomized experimental schedule, and launched the workload in a fresh JVM. The selected collector was enabled using the corresponding JVM option: \texttt{-XX:+UseSerialGC}, \texttt{-XX:+UseParallelGC}, or \texttt{-XX:+UseG1GC}. OpenJDK and Oracle~JDK were executed under otherwise comparable experimental conditions. \\

\noindent Because all three subjects were persistent or interaction-driven rather than terminating programs, each session required a bounded measurement window. Each application was executed for a fixed duration corresponding to its light, medium, or heavy configuration, while scripted API or interface interactions generated the workload. EnergiBridge measurement was started immediately before each session and stopped when the configured duration elapsed. Runs were conducted under controlled system conditions, with non-essential activity limited and pauses between executions to reduce thermal carry-over. \\

\noindent A mock energy interface was also implemented to support pipeline development on systems without access to Linux RAPL counters. This interface was used only to test experiment orchestration and output handling. Measurements produced by the mock interface were not included in the empirical dataset.

\subsection{Energy and Performance Measurement}

Energy and execution-time measurements were collected on the measurement laptop using EnergiBridge~\cite{energibridge}. EnergiBridge reads the Running Average Power Limit counters exposed by the processor and records processor-package energy consumption during program execution. Each run generated an individual EnergiBridge CSV file containing the recorded measurements. The run configuration, execution status, and resulting measurements were subsequently consolidated into a common dataset for statistical analysis. Processor-package energy consumption, measured in joules, was the principal outcome variable. Execution time, measured in seconds, was retained as the primary performance measure. Mean power was derived as the ratio between measured energy and execution time:

\[
P_{\mathrm{mean}} = \frac{E}{t},
\]

where \(E\) denotes processor-package energy consumption in joules and \(t\) denotes execution time in seconds. Energy and execution time were analysed together because lower power demand does not necessarily produce lower total energy consumption when it is accompanied by a longer execution time.

\subsection{Data Preparation and Statistical Analysis}
\label{sec:analysis}

Run metadata and EnergiBridge measurements were consolidated into a common dataset for analysis. A data-integrity check performed prior to modelling identified that, for all runs drawn from the originally planned benchmark subjects, the recorded energy value was numerically consistent with a Unix millisecond timestamp rather than a physically plausible energy reading; the values decoded directly to the date and time at which each run was executed. Because this corruption affected every run in the benchmark subject families and no verified raw energy readings could be recovered for them, those subjects were excluded from analysis in their entirety. The three application subjects (Section~\ref{sec:subjects}) showed no evidence of this or any comparable irregularity: energy values across GC, workload, and JDK conditions were of a magnitude and variability consistent with physical measurement, and all 216 planned runs for these subjects were retained. \\

\noindent The primary outcomes were processor-package energy consumption and execution time. Mean power was derived for each run as \(E/t\), where \(E\) is measured energy consumption and \(t\) is execution time. Descriptive summaries and graphical checks were first used to examine the distributions of the outcomes, identify potential measurement anomalies, and compare the scale of variation within and between subjects. Because only three Java subjects remained after the exclusion described above, subject was treated as a fixed blocking factor rather than a random effect: a random-intercept variance component is not reliably estimable from three groups, and standard practice for RCBD designs at this scale is to model blocks as fixed. Inferential analysis was therefore conducted using a full-factorial RCBD analysis of variance (ANOVA), with garbage collector, workload intensity, JDK distribution, subject, and the garbage-collector-by-workload and garbage-collector-by-JDK interactions entered as fixed effects. \\

\noindent For energy consumption, the model can be expressed as:

\[
\begin{aligned}
E_{ijkr}
={}&
\beta_0
+\beta_{\mathrm{Subject},r}
+\beta_{\mathrm{GC},i}
+\beta_{\mathrm{Workload},j}
+\beta_{\mathrm{JDK},k}
\\
&+\beta_{\mathrm{GC}\times\mathrm{Workload},ij}
+\beta_{\mathrm{GC}\times\mathrm{JDK},ik}
+\varepsilon_{ijkr}.
\end{aligned}
\]

where \(\beta_{\mathrm{Subject},r}\) represents the fixed effect of block \(r\) and
\(\varepsilon_{ijkr}\) represents residual variation. An analogous model was used for the Energy--Delay Product. \\

\noindent Model assumptions were evaluated through residual diagnostics and tests of distributional shape and variance homogeneity. Where the omnibus ANOVA justified pairwise comparisons among collectors or workload levels, Tukey-adjusted post-hoc contrasts were used. Statistical tests used a significance level of \(\alpha=0.05\), and statistical significance was interpreted together with estimated effect sizes because a detectable difference in energy consumption does not necessarily constitute a practically meaningful optimization.

\section{Results}
\label{sec:results}

This section reports the observed energy and performance patterns across garbage collectors, workload intensities, and JDK distributions, restricted to the three verified application subjects (Section~\ref{sec:analysis}). Because only three subjects remain as blocks, the inferential analysis reported here uses a fixed-block RCBD ANOVA rather than a mixed-effects model with subject as a random intercept: a random-intercept variance component is not reliably estimable from three groups, and preliminary mixed-effects fits produced unstable interaction tests as a result. Treating subject as a fixed block is standard practice for RCBD designs and is used throughout this section. The results distinguish descriptive differences from statistically supported effects.

\subsection{Energy Consumption Across Garbage Collectors}
\label{sec:results-gc}

Across the collected measurements, processor energy consumption ranged from approximately 238~J to 3959~J, with a mean of 888.8~J and a standard deviation of 773.8~J. This wide dispersion reflects the heterogeneous resource demands of the evaluated applications and workload intensities. Table~\ref{tab:gc-descriptive} summarizes the descriptive results by collector. Parallel had the lowest mean energy consumption at 839.8~J, followed by Serial at 857.6~J and G1 at 969.0~J. G1 also exhibited the highest mean power, whereas mean execution time was fixed by the bounded measurement window and did not differ meaningfully among collectors. \\

\begin{table}[t]
\centering
\caption{Descriptive results by garbage collector (service-application subset, $n=216$).}
\label{tab:gc-descriptive}
\small
\begin{tabular}{lrrr}
\toprule
\textbf{Collector} & \textbf{Energy (J)} &
\textbf{Runtime (s)} & \textbf{Power (W)} \\
\midrule
Serial   & 857.6 & 200.0 & 4.49 \\
Parallel & 839.8 & 200.0 & 4.31 \\
G1       & 969.0 & 200.0 & 4.96 \\
\bottomrule
\end{tabular}
\end{table}

\begin{figure}[t]
    \centering
    \includegraphics[width=\linewidth]{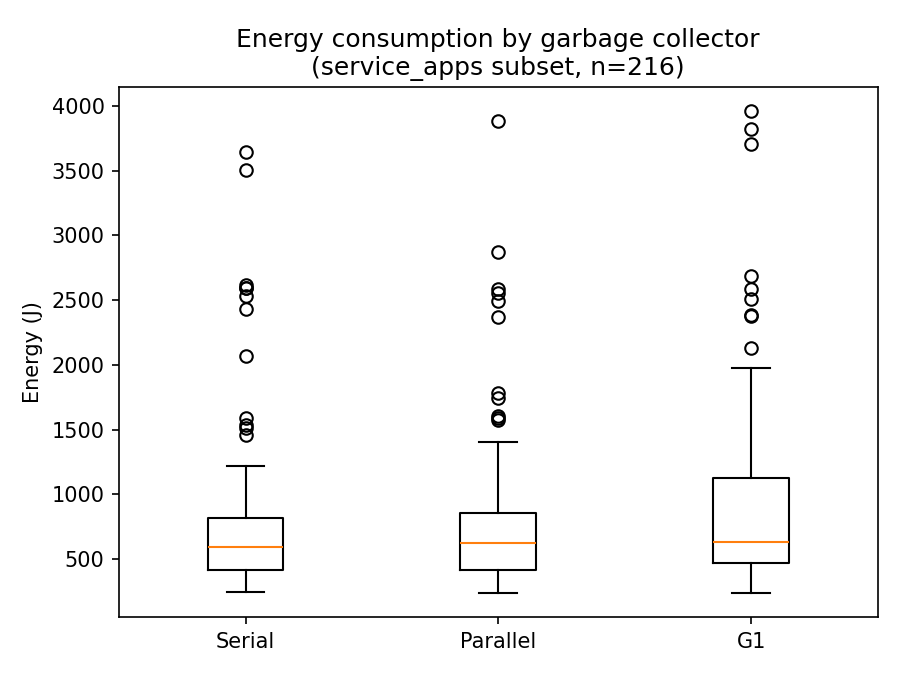}
    \caption{Energy consumption by garbage-collection strategy (service-application subset, $n=216$). Parallel has the lowest descriptive mean, followed closely by Serial and then G1, but the distributions overlap substantially.}
    \label{fig:energy-gc}
\end{figure}

\noindent Figure~\ref{fig:energy-gc} shows that this ordering was visible in the raw distributions, but also that the distributions overlapped substantially. The RCBD ANOVA found no statistically significant main effect of garbage collector (\(F(2,202)=0.65,\ p=0.524\)), and pairwise Tukey-adjusted comparisons among all three collectors were non-significant (all \(p>0.57\)). The lower mean observed for Parallel should therefore be interpreted as a descriptive tendency rather than evidence that any collector was consistently more energy efficient across the three evaluated applications.

\begin{figure}[t]
    \centering
    \includegraphics[width=\linewidth]{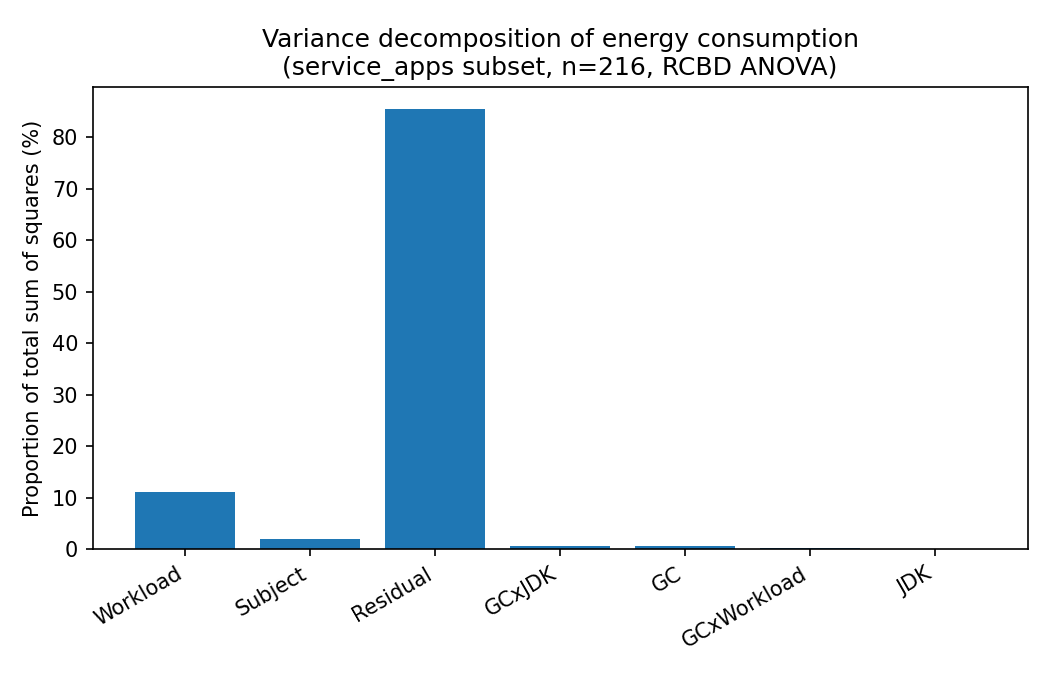}
    \caption{Variance decomposition of energy consumption by experimental factor, expressed as a proportion of total sum of squares (service-application subset, $n=216$).}
    \label{fig:variance-decomposition}
\end{figure}

\noindent Figure~\ref{fig:variance-decomposition} presents the proportion of total sum of squares attributable to each experimental factor in the RCBD ANOVA. Workload accounts for the largest identifiable share of variance (11.1\%), followed by subject (1.9\%) and the GC-by-JDK interaction (0.6\%); garbage collector alone accounts for 0.6\% and its interaction with workload for 0.3\%. The residual (85.4\%) reflects the substantial within-cell variability characteristic of energy measurements on real applications. This decomposition is descriptive; the ANOVA $F$-tests reported above and below remain the basis for inferential claims.

\subsection{Workload Intensity and Collector Interaction}
\label{sec:results-workload}

\begin{figure}[t]
    \centering
    \includegraphics[width=\linewidth]{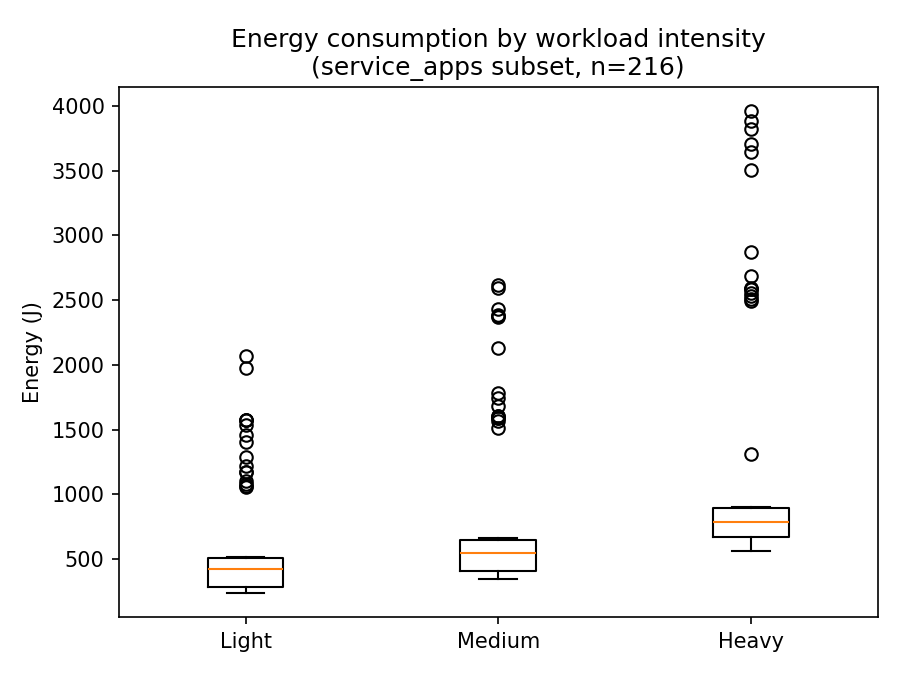}
    \caption{Energy consumption by workload intensity (service-application subset, $n=216$). Heavy differs significantly from Light and Medium; Light and Medium do not differ significantly from each other.}
    \label{fig:workload-energy}
\end{figure}

\noindent Mean energy increased monotonically across the three workload levels: light configurations consumed 606.7~J, medium configurations 830.0~J, and heavy configurations 1229.7~J. The overall workload effect was statistically significant (\(F(2,202)=13.18,\ p<0.001\)). Tukey-adjusted pairwise comparisons showed that Heavy differed significantly from both Light (\(p<0.001\)) and Medium (\(p=0.0036\)), while Light and Medium did not differ significantly from one another (\(p=0.163\)). As shown in Figure~\ref{fig:workload-energy}, workload intensity is the clearest source of systematic variation identified in this experiment. \\

\begin{figure}[t]
    \centering
    \includegraphics[width=\linewidth]{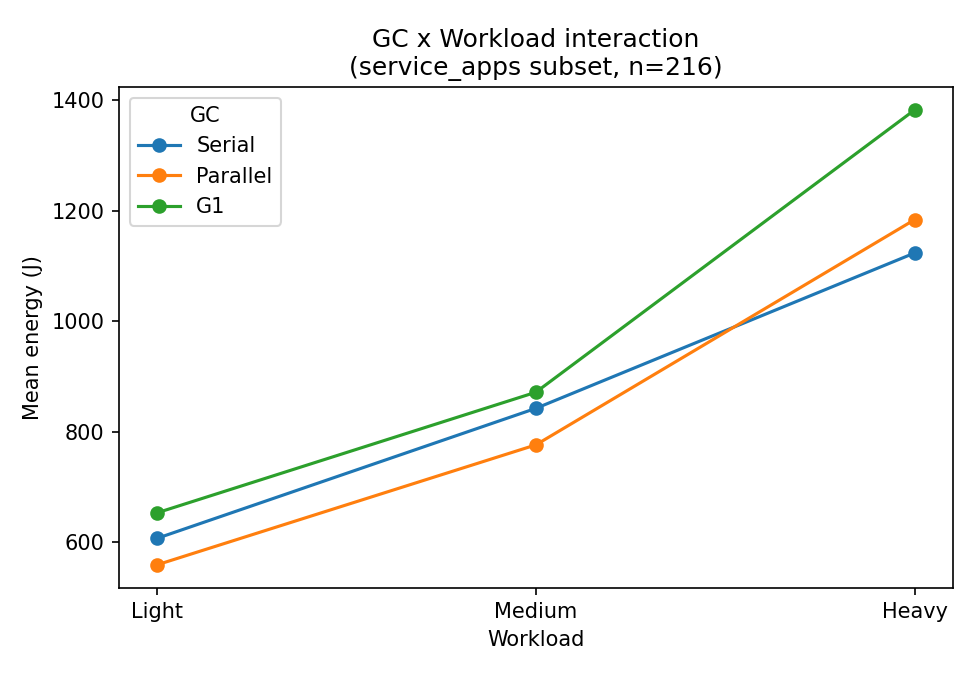}
    \caption{Interaction between garbage collector and workload intensity (service-application subset, $n=216$). Energy increases with workload for all three collectors; the interaction is not statistically significant.}
    \label{fig:gc-workload}
\end{figure}

\noindent The collector-by-workload interaction was not statistically significant (\(F(4,202)=0.18,\ p=0.948\)). Figure~\ref{fig:gc-workload} shows broadly parallel trajectories across workload levels for all three collectors, providing no reliable evidence that increasing workload intensity changed the relative behaviour of the collectors, although G1 shows a numerically steeper descriptive increase from Light to Heavy than Serial or Parallel. \\

\subsection{Energy--Performance Relationship}
\label{sec:results-performance}

Energy consumption and execution time were positively correlated overall (\(r=0.33,\ p<0.001\)), a moderate association rather than the near-proportional relationship that would be expected if runtime alone drove energy demand. The relationship was of comparable magnitude within each collector: \(r=0.28\) for Serial, \(r=0.37\) for Parallel, and \(r=0.36\) for G1 (all \(p<0.02\)). This indicates that longer-running configurations tended toward higher energy use, but the association leaves most of the variation in energy consumption unexplained by runtime alone, most likely reflecting the influence of workload intensity and application-specific allocation behaviour. \\

\begin{figure}[t]
    \centering
    \includegraphics[width=\linewidth]{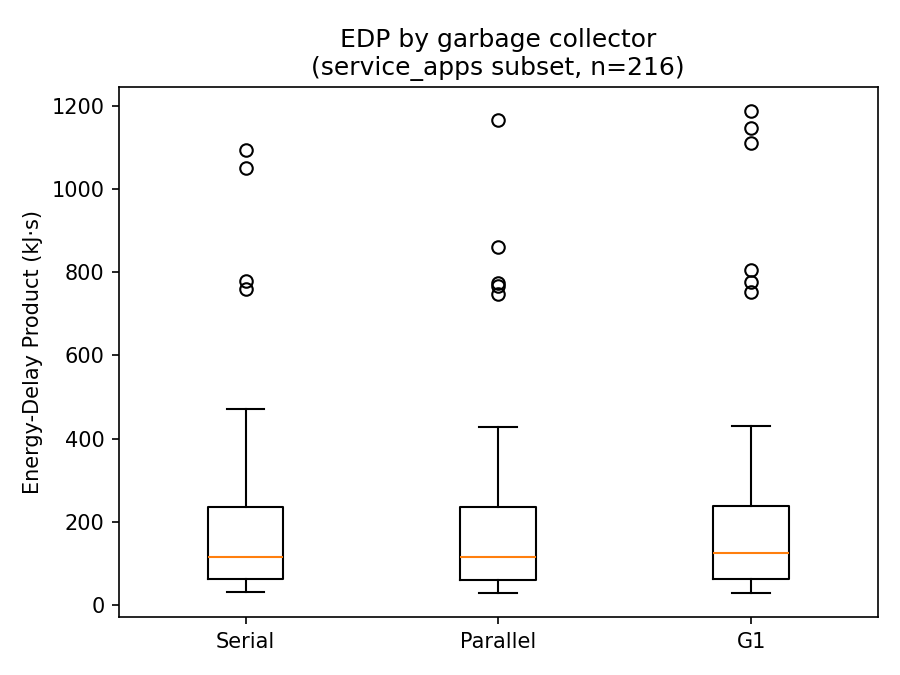}
    \caption{Energy--Delay Product by garbage collector (service-application subset, $n=216$). G1 has a numerically higher descriptive mean, but the collector effect is not statistically significant.}
    \label{fig:edp}
\end{figure}

\noindent Energy--Delay Product (EDP) was examined as a joint energy--performance measure. Mean EDP was similar for Serial (\(187.2\times10^{3}\)~J\,s) and Parallel (\(187.3\times10^{3}\)~J\,s), and higher for G1 (\(216.6\times10^{3}\)~J\,s). An RCBD ANOVA on EDP found no significant effect of collector (\(F(2,205)=0.59,\ p=0.557\)) or of the collector-by-workload interaction (\(F(4,205)=0.30,\ p=0.875\)), while workload intensity remained highly significant (\(F(2,205)=48.1,\ p<0.001\)), consistent with the energy results above. As shown in Figure~\ref{fig:edp}, EDP therefore mirrors the raw-energy analysis: a numerically higher mean for G1, but no statistically reliable collector effect.

\subsection{JDK Distribution}
\label{sec:results-jdk}

The RCBD ANOVA found no statistically significant difference in energy consumption between OpenJDK (863.2~J) and Oracle~JDK (914.3~J) (\(F(1,202)=0.26,\ p=0.611\)). The collector-by-JDK interaction was also non-significant (\(F(2,202)=0.71,\ p=0.494\)). Table~\ref{tab:jdk-gc} reports the descriptive marginal means; although Parallel appears notably lower under OpenJDK than under Oracle~JDK, this pattern is not statistically supported by the interaction test and should be treated as sampling variation given the modest cell sizes (36 runs per JDK-by-collector cell) rather than as evidence of JDK-dependent collector behaviour. As with the collector main effect, the absence of a statistically detectable JDK difference should not be interpreted as evidence that OpenJDK and Oracle~JDK are equivalent in energy consumption; no formal equivalence test was performed on this subset.

\begin{table}[t]
\centering
\caption{Mean energy by JDK distribution and garbage collector (service-application subset, descriptive only; $n=36$ per cell).}
\label{tab:jdk-gc}
\small
\begin{tabular}{llr}
\toprule
\textbf{JDK} & \textbf{Collector} & \textbf{Energy (J)} \\
\midrule
OpenJDK & Serial   & 909.0 \\
OpenJDK & Parallel & 745.5 \\
OpenJDK & G1       & 935.2 \\
Oracle  & Serial   & 806.3 \\
Oracle  & Parallel & 934.0 \\
Oracle  & G1       & 1002.8 \\
\bottomrule
\end{tabular}
\end{table}

\subsection{Summary of Findings}
\label{sec:results-summary}

\noindent Three findings emerge from the analysis. First, Parallel and Serial had lower descriptive mean energy consumption than G1, but the RCBD ANOVA did not establish a statistically reliable collector effect, and this held for both raw energy and EDP. Second, workload intensity was the only factor to reach statistical significance, with Heavy workloads consuming significantly more energy than Light or Medium; the collector-by-workload interaction was not significant, so this effect did not depend on which collector was used. Third, execution time and energy showed a moderate, not near-proportional, positive association across all collectors, and JDK distribution produced neither a detectable main effect nor an interaction with collector. Taken together, the results do not identify a universally energy-optimal garbage collector or a reliable difference between the two evaluated JDK distributions within this reduced, three-application dataset. Collector selection may still matter for individual applications, as suggested by the numerically higher G1 means, but the direction and magnitude of that effect cannot be generalized from the aggregate evidence available here.

\section{Threats to Validity}
\label{sec:threats}

This study evaluates garbage-collector behaviour under a controlled experimental
configuration, but its findings remain subject to several limitations. These limitations are discussed in terms of internal, construct, external, and conclusion
validity.

\subsection{Internal Validity}

Energy measurements may have been influenced by background operating-system activity, thermal variation, CPU scheduling, and other processes unrelated to the experimental treatments. To limit such interference, the device under test was dedicated to workload execution, non essential activity was restricted, and orchestration was performed remotely on a Raspberry~Pi. Runs were separated by cooldown periods and executed in randomized order so that transient environmental effects were less likely to align systematically with a particular garbage collector, workload, or JDK distribution. These controls reduce, but cannot
eliminate, measurement noise. \\

\noindent JVM initialization and JIT compilation may also influence runtime and energy, particularly during short executions. Workloads were executed in fresh JVM processes, and warm-up execution was used before measurement where applicable. Nevertheless, collector heuristics and compiled code can continue to evolve during execution, meaning that not every observation necessarily represents a fully stabilized JVM state. Although the experimental factors were controlled, the study was performed on one
physical device. Repeated runs therefore represent replications of configurations
on the same system rather than independent replications across machines. Any
unobserved device-specific behaviour may consequently affect all measurements.

\subsection{Construct Validity}

Energy consumption was operationalized using the processor energy counters exposed
through the Linux powercap interface and collected by \texttt{EnergiBridge}~\cite{energibridge}. These measurements represent the energy domains made available by the tested AMD system; they do not capture the complete wall-socket energy consumption of the laptop. Energy used by components outside the exposed domains, including the display, storage, networking hardware, and power-conversion circuitry, may therefore be omitted. The results should accordingly be interpreted as processor-level energy estimates rather than total system energy consumption. \\

\noindent Runtime and Energy--Delay Product provide complementary views of performance and
its relationship with energy, but neither measure captures every aspect of
application quality. Throughput, response-time distributions, memory pressure, and
service-level requirements may alter whether a collector is preferable in a
production deployment. The treatment space was also limited to Serial, Parallel, and G1 garbage
collection. These collectors represent established JVM strategies with different
design goals, but they do not cover the complete range of available collectors. In particular, the results do not characterize low-latency collectors such as ZGC or Shenandoah.

\subsection{External Validity}

All measurements were collected from a single AMD Ryzen~9~8940HX laptop running
Ubuntu Linux and the selected OpenJDK and Oracle JDK configurations. The results
may differ on server-class processors, ARM systems, virtual machines, cloud
instances, other operating systems, or machines with different memory and thermal
characteristics. The findings should therefore not be generalized directly beyond
hardware and software environments comparable to the evaluated testbed. \\

\noindent The experiment evaluated three Java applications: Spring PetClinic, a REST-based To-Do application, and the ANDIE image-processing tool, each exercised under light, medium, and heavy configurations. This is a small subject pool for an RCBD design, and it is narrower than originally planned: a parallel set of terminating benchmark subjects (DaCapo, the Computer Language Benchmarks Game, and Rosetta Code) was excluded after a data-integrity check found that their recorded energy values were corrupted at the source (Section~\ref{sec:analysis}), rather than for reasons related to their behaviour under test. The three retained applications share a persistent, interaction-driven execution style; the results therefore speak most directly to server-side Java services and should not be assumed to generalize to short-lived, terminating, or batch-style Java programs, which were not represented in the verified dataset. The workload labels are relative to the configurations used in this study and should not be interpreted as standardized categories applicable to every Java system. Large heaps, highly concurrent applications, and latency-sensitive workloads outside the three evaluated applications may produce different collector behaviour. \\

\noindent The application workloads were executed through bounded, scripted sessions.
Although this made them measurable within the common experimental design, such
sessions approximate rather than reproduce long-running production operation.
They may not capture effects such as heap ageing, changing request distributions,
traffic bursts, or collector adaptation over extended periods.

\subsection{Conclusion Validity}

Of the 216 runs planned for the three verified application subjects, all 216 passed data-quality screening and were retained for analysis, indicating that the corruption affecting the excluded benchmark subjects did not extend to this subset. Because the exclusion was systemic to entire subject families rather than scattered across individual runs, it is unlikely to have introduced the kind of condition-specific bias that partial, run-level exclusions would raise; however, it substantially reduces the number of independent blocks available to the design, from six subject families to three applications. With only three blocks, subject is treated as a fixed rather than random effect in the RCBD ANOVA (Section~\ref{sec:analysis}), which is standard practice at this scale but means that between-subject variability is estimated, not generalized to a wider population of possible Java applications. \\

\noindent Energy measurements displayed substantial within-condition variability and broad overlap among collectors. Consequently, the absence of statistically significant
collector or JDK effects should not be interpreted as proof that the corresponding
effects are zero. Small or subject-specific differences may remain undetected
because of measurement noise, heterogeneous workloads, and the limited number of
subjects. Model diagnostics were inspected to assess residual behaviour and the influence of unusual observations, and the variance decomposition reported in Section~\ref{sec:results-gc} is derived directly from the same ANOVA sum-of-squares used for the significance tests, so the two are not expected to diverge in the way that an exploratory decomposition and a separately fitted model can. \\

\noindent Finally, multiple descriptive comparisons were examined across collectors,
workloads, JDKs, and combined configurations. Configuration-level patterns should
therefore be treated as exploratory unless supported by the corresponding
inferential tests. In particular, non-significant mean differences do not justify
a universal ranking of the evaluated collectors, and the reduced subject pool means these conclusions should be read as applying to the three evaluated applications rather than to Java server-side applications in general.

\section{Discussion}
\label{sec:discussion}

This study examined how garbage-collector strategy, workload intensity, and JDK distribution relate to processor-level energy consumption and execution time across three Java applications. The results present a more qualified picture than a simple ranking of collectors. Parallel and Serial exhibited lower descriptive mean energy consumption than G1, but the differences were small relative to the substantial variability within each collector and were not statistically significant. Workload intensity, by contrast, was the one factor to produce a clear, statistically supported effect. Execution time and energy showed a positive but moderate association, weaker than a near-proportional relationship would imply.

\subsection{Interpretation of the Findings}

\paragraph{Collector differences were visible but statistically uncertain.}

Parallel recorded the lowest descriptive mean energy consumption (839.8~J), followed closely by Serial (857.6~J) and G1 (969.0~J). G1 therefore consumed approximately 15\% more energy than Parallel and 13\% more than Serial at the aggregate descriptive level. The corresponding plots showed a broadly similar ordering across several configurations. However, the energy distributions overlapped substantially, and the RCBD ANOVA did not establish a statistically significant collector effect (\(F(2,202)=0.65,\ p=0.524\)), a result that held for both raw energy and Energy--Delay Product. The observed ranking therefore does not support the conclusion that Parallel or Serial is universally more energy-efficient than G1. Instead, it suggests a possible aggregate tendency whose reliability could not be distinguished from subject-level variation and measurement noise across the three evaluated applications. The variance decomposition in Section~\ref{sec:results-gc} attributes only a small share of total variance to collector (0.6\%) relative to workload (11.1\%) and residual variation (85.4\%), consistent with the ANOVA result rather than in tension with it.

\paragraph{Workload intensity was the clearest statistically supported finding.}

Mean energy increased from 606.7~J under light workloads to 830.0~J under medium workloads and 1229.7~J under heavy workloads. This monotonic pattern is consistent with the expectation that more demanding configurations require greater computation and therefore more energy, and unlike the collector effect, it was statistically significant (\(F(2,202)=13.18,\ p<0.001\)). Post-hoc comparisons localized this effect to the Heavy condition: Heavy differed significantly from both Light and Medium, while Light and Medium did not differ significantly from each other. The collector-by-workload interaction was not significant, so this effect did not depend on which collector was in use; G1 showed a numerically steeper descriptive increase from Light to Heavy than Serial or Parallel, but the interaction test does not support treating this as a reliable collector-specific sensitivity to workload. Workload intensity is therefore the strongest lever identified in this experiment for managing energy consumption, and it operates independently of collector choice.

\paragraph{Energy and runtime were positively but only moderately associated.}

The observed relationship between execution time and energy consumption was positive and statistically significant but moderate in strength (\(r=0.33,\ p<0.001\)), with comparable correlations within Serial (\(r=0.28\)), Parallel (\(r=0.37\)), and G1 (\(r=0.36\)). Configurations that ran longer tended to consume more energy, but runtime alone accounts for only a modest share of the variation in energy consumption; workload intensity and application-specific allocation behaviour evidently contribute substantially more. \\

\noindent This association should not be interpreted as evidence that execution time alone caused the observed energy differences. Both runtime and energy can increase because a workload performs more computation, processes more data, or places greater pressure on memory and the collector. The finding indicates a real but partial energy--performance alignment: reducing unnecessary execution time is plausibly beneficial for energy use, but it is not a reliable proxy for energy consumption on its own. The Energy--Delay Product analysis reinforced this picture. Serial and Parallel had similar, lower descriptive EDP than G1, but neither collector, workload, nor their interaction produced a statistically significant collector-related effect on EDP beyond the workload main effect already noted above. The experiment therefore did not identify a collector-attributable energy--performance trade-off, though it also cannot rule one out in individual applications outside the evaluated set.

\paragraph{No JDK difference was established.}

OpenJDK and Oracle~JDK produced similar descriptive energy distributions (863.2~J and 914.3~J respectively), and neither the JDK main effect (\(p=0.611\)) nor the collector-by-JDK interaction (\(p=0.494\)) was statistically significant. Collector behaviour therefore appeared broadly stable across the two evaluated distributions. As with the collector effect, non-significance here does not demonstrate that the distributions are equivalent; no formal equivalence test was performed on this dataset. The data available do not support choosing between these distributions on the basis of processor-level energy consumption alone.

\subsection{Practical Implications}

For application developers and operators, the findings argue against choosing a collector solely from aggregate energy rankings. Parallel and Serial had lower descriptive means than G1, but the uncertainty around these differences prevents a general recommendation. Collector selection should remain application-specific and account for throughput, latency, pause-time, memory-footprint, and stability requirements in addition to energy. By contrast, workload intensity was a reliable and substantial driver of energy consumption in this experiment, making workload-aware capacity planning a more evidence-backed lever than collector substitution for the applications studied here. The moderate energy--runtime association means performance profiling is a reasonable starting point for energy work, but runtime should not be treated as a complete substitute for direct energy measurement, since configurations with similar execution times can still differ in energy demand. \\

\noindent The absence of a detectable JDK effect simplifies the choice between the two tested distributions only in a limited sense: energy measurements provide no clear reason to prefer one over the other in this testbed. Compatibility, support, licensing, operational tooling, and application performance remain independent considerations. More broadly, the results favour measurement-guided configuration over universal tuning rules. The collector with the lowest aggregate mean need not be the best option for every application, workload, or deployment environment. Teams seeking energy reductions should therefore benchmark realistic application behaviour on hardware comparable to their production systems, and should treat workload management, not just collector choice, as a primary energy lever.

\subsection{Implications for Sustainable Java Systems}

The findings show that energy efficiency in Java cannot be reduced to a single JVM flag. Collector choice produced observable descriptive differences, but those differences were not sufficiently consistent to establish a general collector effect, even as workload intensity emerged as a clear and substantial driver of energy demand. For sustainable software engineering, this shifts attention from searching for a universally ``green'' collector toward understanding how workload characteristics, execution time, allocation behaviour, and collector policy interact. JVM configuration remains relevant, but the evidence here places workload management ahead of it as a practical energy-management priority; collector choice should be evaluated as one component of a broader system rather than as an independent solution to energy inefficiency. \\

\noindent Future studies should examine this interaction using a larger and more diverse set of subjects than was available after data-integrity screening in this experiment, additional hardware platforms, and collectors such as ZGC and Shenandoah. Direct wall-power measurements would complement the processor-level energy counters used here, while longer-running service experiments could reveal effects related to heap ageing, adaptive compilation, traffic variation, and latency that bounded sessions may not capture.

\bibliographystyle{IEEEtran}
\bibliography{references}

\end{document}